\documentclass[12pt,a4paper]{article}
\usepackage[left=1in,right=1in,top=1.5in,bottom=1.5in]{geometry}

\usepackage{mathrsfs}
\usepackage{amssymb}
\usepackage{amsmath}
\usepackage{ascmac}
\usepackage{amsthm}
\usepackage[dvipdfmx]{graphicx}
\usepackage{booktabs}
\usepackage{enumerate}
\usepackage{setspace}
\usepackage{natbib}
\usepackage{color}
\usepackage{xcolor}
\usepackage{url}
\usepackage{comment}
\usepackage[colorlinks,linkcolor=blue,citecolor=blue,urlcolor=blue]{hyperref} 
\usepackage{times}
\usepackage{xcolor}

\usepackage{algpseudocode}
\usepackage{algorithm}
\algnewcommand{\Indent}[1]{\State\hspace{#1}}

\usepackage{titlesec}
\titleformat*{\section}{\large\bfseries}
\titleformat*{\subsection}{\it}

\usepackage[english]{babel} 
\usepackage{blindtext}
\usepackage{enumitem}
\usepackage{indentfirst}
\usepackage{graphicx}
\usepackage{tabularx}
\usepackage{caption}

\newcommand{\revise}[1]{\textcolor{black}{#1}}

\title{{\bf Clustered Factor Analysis for Multivariate \\ 
Spatial Data}\footnote{This version: \today}}
\date{}

\begin{document}
\doublespacing
\maketitle

\vspace{-2cm}
\begin{center}
{\large Yanxiu Jin$^1$, Tomoya Wakayama$^2$, Renhe Jiang$^1$ and Shonosuke Sugasawa$^3$ 
}
\end{center}

\noindent
$^1$Center for Spatial Information Science, The University of Tokyo\\
$^2$Graduate School of Economics, The University of Tokyo\\
$^3$Faculty of Economics, Keio University

\vspace{0.5cm}
\begin{center}
{\large\bf Abstract}
\end{center}

Factor analysis has been extensively used to reveal the dependence structures among multivariate variables, offering valuable insight in various fields. However, it cannot incorporate the spatial heterogeneity that is typically present in spatial data. 
To address this issue, we introduce an effective method specifically designed to discover the potential dependence structures in multivariate spatial data. Our approach assumes that spatial locations can be approximately divided into a finite number of clusters, with locations within the same cluster sharing similar dependence structures. By leveraging an iterative algorithm that combines spatial clustering with factor analysis, we simultaneously detect spatial clusters and estimate a unique factor model for each cluster. 
The proposed method is evaluated through comprehensive simulation studies, demonstrating its flexibility. In addition, we apply the proposed method to a dataset of railway station attributes in the Tokyo metropolitan area, highlighting its practical applicability and effectiveness in uncovering complex spatial dependencies.


\section{Introduction}

Multivariate statistical analysis refers to a range of statistical methods aimed at investigating the dependency structure among multiple variables to understand and explain complex datasets. These approaches are particularly valuable in disciplines where the interactions between variables are intricate and potentially influenced by spatial relationships, such as environmental science \citep{b2020multivariate}, sociology \citep{morales2022gender}, epidemiology \citep{Eric2021}, and urban planning \citep{SONG2020102072}. As the number of variables increases, statistical model performance and data visualization face challenges. Data dimensionality reduction techniques address these challenges by simplifying the dataset while retaining as much important information as possible. These techniques reduce the number of variables either by feature selection \citep{abdulwahab2022feature,hancer2020survey,khaire2022stability} or feature extraction \citep{fu2020fusion,shrestha2021factor}. Factor analysis is one representative method for feature extraction. It identifies a few underlying factors that explain the patterns of correlations among observed variables. These factors are defined as linear combinations of the original variables for ease of interpretation and computation. 

The exploratory factor analysis (EFA) model excels at reducing data dimensions by assuming that the relationships between observed variables are homogeneous ~\citep{johnson2002applied,fabrigar2012exploratory}. However, this assumption often has flaws in geographic data, \revise{where spatial heterogeneity plays a significant role. As stated in the ``Second Law of Geography" \citep{axinghowisthe2022}, spatial heterogeneity implies that geographic variables exhibit uncontrolled variance across space, leading to location-specific variations. This means that the relationships between variables can vary considerably depending on their locations.} Hence, when EFA is performed on multivariate spatial data, it does not capture the spatial heterogeneity of correlations between variables. 

Some studies have attempted to consider the factor model under spatial correlation \citep{dey2022graphical,wang2003generalized,krupskii2018factor}. For instance, \cite{wang2003generalized} assumed that a common spatial factor influences observed variables at different locations, and used Bayesian methods and Markov chain Monte Carlo computational techniques to estimate parameters and predict the common spatial factor. Although these studies effectively simulate how spatial proximity influences the observed values, they may not accurately distinguish how locations inhomogeneously affect the relationships between variables. This could lead to a neglect of spatial heterogeneity, where significant differences in variable relationships observed between different geographical locations may exist. Therefore, this limitation may restrict the applicability of the models, especially to diverse or heterogeneous structural environments.

Another way to consider spatial attributes is to incorporate spatial heterogeneity into the modeling process, allowing regression coefficients to vary spatially. This includes widely used techniques in spatial modeling such as geographically weighted regression~\citep{fotheringham2003geographically}, spatially clustered regression~\citep{sugasawa2021spatially}, and spatial cluster detection~\citep{lee2017cluster}. The advantage of these models lies in their ability not only to handle the spatial correlation between variables but also to reveal the varying strength of relationships between different geographical regions. Although these existing methods deal with spatial heterogeneity, they mainly focus on standard regression problems and have not yet achieved the interpretable dimensionality reduction regression that is the benefit of EFA.


To overcome this issue, by introducing clustering methods and geographic weights into factor models, we develop spatially clustered factor analysis (SCFA) that combines the treatment of spatial heterogeneity with the factor model framework.
We assume that the samples can be divided into a finite number of groups and that the underlying structure of the observed variables in the same groups is consistent. First, we forcibly divide samples into several groups, where grouping rules can follow geographical proximity or be randomly assigned. Next, to allow for spatial variation in the SCFA, we use indicators representing the group to which each location belongs and simultaneously estimate both group parameters and factor models. Additionally, to encourage such spatial clustered structure, we incorporate a penalized likelihood function proposed by \cite{sugasawa2021spatially}, which is based on the hidden Potts model \citep{potts1952some}. \revise{Similar to the Mixture of Factor Analyzers (MFA) model proposed by \cite{ghahramani1996algorithm}, SCFA employs an expectation-maximization (EM) algorithm, iteratively optimizing group assignment and factor model parameters until convergence. However, unlike MFA, which does not consider spatial information, SCFA explicitly integrates spatial dependencies through geographic weights, enabling it to capture spatial heterogeneity more effectively.} We will demonstrate that the SCFA method can be easily implemented through a simple iterative algorithm similar to the \emph{K-means} algorithm~\citep{macqueen1967some}, which combines the existing algorithms of the EFA with straightforward optimization steps for group assignment. Although grouping methods using indicators (such as the method proposed) have been widely used in panel data analysis \citep{ito2023grouped,wang2018homogeneity} and clustered data \citep{sugasawa2021grouped}, no research has ever applied them to factor analysis of multivariate spatial data as far as we know.

This paper is organized as follows. In Section \ref{sec:2}, we introduce the proposed method, detailing the theoretical foundation and computational framework that underpins our approach. Section \ref{sec:3} presents a comprehensive simulation study to evaluate the performance and accuracy of the method under various spatial scenarios. In Section \ref{sec:4}, we apply the SCFA method to a real-world dataset, showcasing its practical utility in capturing spatial heterogeneity and the relationship between factors and variables. Finally, the contributions of this study and potential future applications are discussed in Section \ref{sec:5}.

\section{Methods}\label{sec:2}
\subsection{Review: Exploratory Factor Analysis}
In EFA, observed variables ($X$) are described as a linear combination of fewer unobservable random variables ($F$). The model equation for the $i$th subject when $m$ unobservable variables are considered for modeling the $p$ $(\ge m)$ observed variables can be written as,
\begin{equation*}
X_i= \sum_{j=1}^m a_{ij}F_j+\varepsilon_i \quad (i=1, 2,\dots, p),
\end{equation*}
where $F_j\  (j=1, 2,\dots, m)$ denotes the common factor, $a_{ij} \ (i=1, 2,\dots, p, j=1, 2,\dots, m)$ is the factor loading of $i$th variable on $j$th factor, $A=(a_{ij})$ is the $p\times m$ matrix of factor loadings, and $\varepsilon_i$ is the specific factor, which is cannot be explained by the $m$ common factors. In compact notation,
\begin{equation*}
    X = AF+\varepsilon,
\end{equation*}
where $X=(X_1,X_2,\dots,X_p)^{\top}$, $F=(F_1,F_2,\dots,F_m)^{\top}$, and $\epsilon=(\varepsilon_1,\varepsilon_2,\dots,\varepsilon_i)^{\top}$.

Under the assumption of the orthogonal factor model \citep{johnson2002applied}, the common factors and specific factors are independent of each other, so that ${\rm Cov}(\varepsilon_i, F_j)=0$; the common factors are independent normal variables with the mean of 0 and the variance of 1, and their covariance matrix is the identity matrix $I_m$, that is, $F \sim \mathcal{N}(0, I_m)$; the special factor $\varepsilon_i\sim \mathcal{N}(0,\sigma_{i}^2)$, and the variances are not necessarily equal with $\Psi=Var(\epsilon)=diag(\sigma_1^2,\sigma_2^2,\cdots,\sigma_p^2)$. Based on these assumptions, the observed variables' variance-covariance matrix $\Sigma$ can be represented as follows:
\begin{equation}
    \Sigma = AA^{\top} + \Psi
    \label{covariance matrix}
\end{equation}

Suppose $X_i \sim \mathcal{N}(\mu,\Sigma)$ is a multivariate normal vector, the unknown parameters $A$ and $\Psi$ can be estimated using the maximum likelihood estimation.
Further, the number of factor $m$ can be determined in a data dependent manner by using, for example, Akaike information criterion (AIC) and Bayesian information criterion (BIC) \citep[e.g.][]{choi2019model}.

\subsection{Spatially Clustered Factor Analysis}\label{subsec:2.2}

Assuming each observed sample $x_i$ has corresponding location information $s_i$ (e.g., longitude and latitude). Then, we divide the $p$ samples into $G$ groups and apply the EFA to each group. The samples belonging to the same group share the same $A_i$ and $\Psi_i$. We consider $G$ as a fixed value for a while, but we will discuss the data-driven selection of $G$ later. We introduce $g_i\in{1,2,\dots, G}$, an unknown group membership variable for the $i$th location, and let $A_i=A_{g_{i}}$ and $\Psi_i=\Psi_{g_{i}}$. Hence, the unknown parameters in the model are the structural parameters $A_i=(A_1, A_2,\dots, A_G)^{\top}$, $\Psi_i=(\Psi_1,\Psi_2,\dots,\Psi_G)^{\top}$, and the membership parameter $g=(g_1,g_2,\dots,g_p)^{\top}$.

When considering the membership parameter, it is reasonable to assume that members in adjacent locations may share the same membership, as the observed data in adjacent locations may exhibit similar features due to common underlying factors. To promote the formation of this structure, we introduce the following penalized likelihood function, motivated by the Potts model \citep{potts1952some} and first adopted by \cite{sugasawa2021spatially}:
\begin{equation}
    Q(A,\Psi, g) \equiv \sum_{i=1}^{p} \log f(x_i | A_{g_{i}},\Psi_{g_{i}}) + \phi \sum_{i<l} w_{il} I(g_i = g_l).
    \label{penalized_likelihood_function}
\end{equation}
This objective function $Q(A,\Psi, g)$ for the maximization problem can be seen as a combination of the logarithm of the joint probability density function and the weight function. The joint probability density function is used to observe the likelihood of samples $x_i$ following a multivariate normal distribution $X_i \sim \mathcal{N}(0, A_{g_{i}}A_{g_{i}}^{\top} + \Psi_{g_{i}})$; the latter term is used to introduce constraints on spatial dependence. $w_{il}=w(s_i,s_l)\in[0,1]$
is a weight function, and $\phi$ is considered a hyper-parameter that regulates the strength of spatial similarity. Here, $w(\cdot,\cdot)$ is a decreasing function of the distance between two points (e.g., $\exp(-\|s_i-s_j\|_2)$), reflecting the idea that the closer the elements are within the same group, the more reasonable it is. 

We adopt an iterative algorithm similar to \emph{K-means} clustering to maximize the objective function \eqref{penalized_likelihood_function}. In this process, it involves continuously updating the membership variable $g$ and the parameters ($A$, $\Psi$) of the factor analysis models. Each step of the algorithm update is very simple: maximizing the objective function under known $g$ conditions is essentially equivalent to maximizing the penalized log-likelihood function for each sample within each group. The detailed steps of the algorithm are shown in Algorithm~\ref{alg:1}. The function $f(X_g, m)$ in Algorithm~\ref{alg:1} refers to the exploratory factor analysis model. To update the membership value $g_i$, we only need to calculate the group $g$ corresponding to the maximum penalty likelihood function for each sample. As long as $G$ is a moderate value, the update process is feasible with low computational intensity. Regarding the condition of convergence, we monitor the difference between the current iteration and the previous iteration:
\begin{equation*}
    D^{(k)}=\sum\limits_{g=1}^{G} \frac{\text{Tr}(|\Psi^{(k)}_g-\Psi^{(k-1)}_g|)}{\text{Tr}(\Psi^{(k-1)}_g)},
\end{equation*}
The algorithm should terminate when the difference $D^{(k)}$ is less than the user-specified tolerance value $\delta$, which we use $\delta = 10^{-6}$ in the simulation study.


\begin{algorithm}[t] 
  \caption{Spatically Clustered Factor Analysis}  \label{alg:1}
  \renewcommand{\algorithmicrequire}{\textbf{Input:}}
  \renewcommand{\algorithmicensure}{\textbf{Output:}}
  \label{alg::CFA}  
  \begin{algorithmic}[0]  

    \State Set initial values $g^{(0)}$, $A^{(0)}$, $\Psi^{(0)}$ and $m$;
    \State \quad - For each $i$th sample, assign initial membership parameter $g^{(0)}_i$   
    \State \quad - For $g=$[1,\dots, $G$], initialize $A^{(0)}_g$, $\Psi^{(0)}_g$ as the \(p \times m\) zero matrix, denoted \(\mathbf{0}_{p \times m}\).
    \State  
    \Repeat  
        \State - Execute exploratory factor analysis for each group to update $A_{g}$, $\Psi_{g}$:    
        \For{$g=$[$1,\dots, G$]}:
        \State $A^{(k)}_g$, $\Psi^{(k)}_g$ = $f(X^{(k)}_g,m)$
        \EndFor
        \State  
        \State - For each $i$th sample, update the membership variable:  
        \State \quad $g^{(k+1)}_i = \underset{g \in \{1, \ldots, G\}}{\arg\max} \left\{ \log f(x_i | A^{(k)}_g; \Psi^{(k)}_g) + \phi \sum\limits_{l=1; \, l \neq i}^{p} w_{il} I(g = g^{(k)}_l) \right\}$.
        \State  
    \State  
    \Until{ convergence}  
  \end{algorithmic}  
\end{algorithm}

\subsection{Selection of tuning parameters}\label{sec:2.3}
In the proposed method, there are three tuning parameters: $G$, representing the number of groups, $m$, the number of common factors, and $\phi$, which controls the strength of spatial dependence of $g_{is}$. Given that the choice of $\phi$ has minimal impact as long as it remains strictly positive, we suggest setting $\phi=1$ for simplicity, as \cite{sugasawa2021spatially} discussed. The number of groups, $G$, can either be predetermined based on prior information about the dataset or determined through a data-driven approach \citep{sugasawa2021spatially} using the following information criterion: 
\begin{equation}
IC(G) = -2 \sum_{i=1}^{n} \log f(x_i \mid \hat{A}_i; \hat{\Psi}_i) + c_nG \left(\text{num}(A) + \text{num}(\Psi)\right),
\label{IG}
\end{equation}
where $c_n$ is a constant depending on the sample size $n$, and \text{num}$(A)$ and \text{num}$(\Psi)$ denote the number of elements in the matrix $A$ and $\Psi$, respectively. Specifically, we use $c_n = \log(n)$, which leads to a BIC-type criterion. Then, we choose a suitable value of $G$ as $\hat{G} = \arg\min_{G \in \{G_1, \ldots, G_L\}} IC(G)$, where $G_1, \ldots, G_L$ are candidates for $G$.

For the number of common factors, $m$, it can be determined by prior information or a data-driven approach, similar to the parameter $G$. In this paper, $m$ is selected based on the parallel analysis introduced by \cite{horn1965rationale}. This is a common way to determine how many factors to extract when performing factor analysis. Unlike the Kaiser criterion \citep{kaiser1960application} of retaining factors with eigenvalues greater than 1, the parallel analysis compares the eigenvalues derived from data with those generated from randomly simulated datasets of the same size and structure. The rationale is that factors whose eigenvalues exceed the eigenvalues from the simulated random data represent true underlying factors in the data, rather than noise. Mathematically, for each factor $m$, if the eigenvalue $\lambda_m$ of the data is greater than the corresponding average or percentile eigenvalue $\lambda_m^{sim}$ of the simulated data sets, the factor is considered significant and retained. This approach helps prevent both over-extraction and under-extraction of factors, balancing accuracy and interpretability.


\section{Simulation Study}\label{sec:3}
\subsection{Simulation settings}

We present simulation studies to illustrate the performance of the SCFA together with non-spatial EFA under six scenarios for different cluster locations. In all scenarios, for $i=1,\ldots,n$, we let $s_i$ be the two-dimensional vector of location information (longitude and latitude) in the squared domain $[-1, 1]^2\subset \mathbb{R}^2$. The spatial locations $\{s_i\}$ are grouped into spatial clusters $D=(D_1, D_2,\dots, D_G)$, with specific definitions provided in the respective scenarios. Recall that the factor model is formulated as follows:
$$
x_i=A_{g_i} f_i + \varepsilon_i, \ \ \ \ \varepsilon_i\sim N(0, \sigma^2 I_p),
$$
where $f_i\sim N(0, I_m)$ with $m<p$, and $A_i$ is a $p\times m$ coefficient matrix.
Here $g_i$ denotes the group membership, that is, $g_i=g$ if $i\in D_g$. 
We set $n=200$, $p=10$, $m=3$ and $G=4$ in our study. 
Let $a_{gjk}$ be the $(j,k)$-element of $D_g$, and ($a_{g1k},\ldots,a_{gpk}$) (the $k$th column vector of $A_g$) are generated from $N(\mu_{gk}, \tau^2_g)$ for $g=1,\ldots,4$, where 
\begin{align*}
&(\mu_{11},\mu_{12},\mu_{13})=(1,1,1), \ \ \ 
(\mu_{21},\mu_{22},\mu_{23})=(-1,0.5, 0.5), \\
&(\mu_{31},\mu_{32},\mu_{33})=(0.5,-1,0.5), \ \ \ 
(\mu_{41},\mu_{42},\mu_{43})=(0.5,0.5,-1).
\end{align*}

\begin{figure}[t]
    \includegraphics[width=1\linewidth]{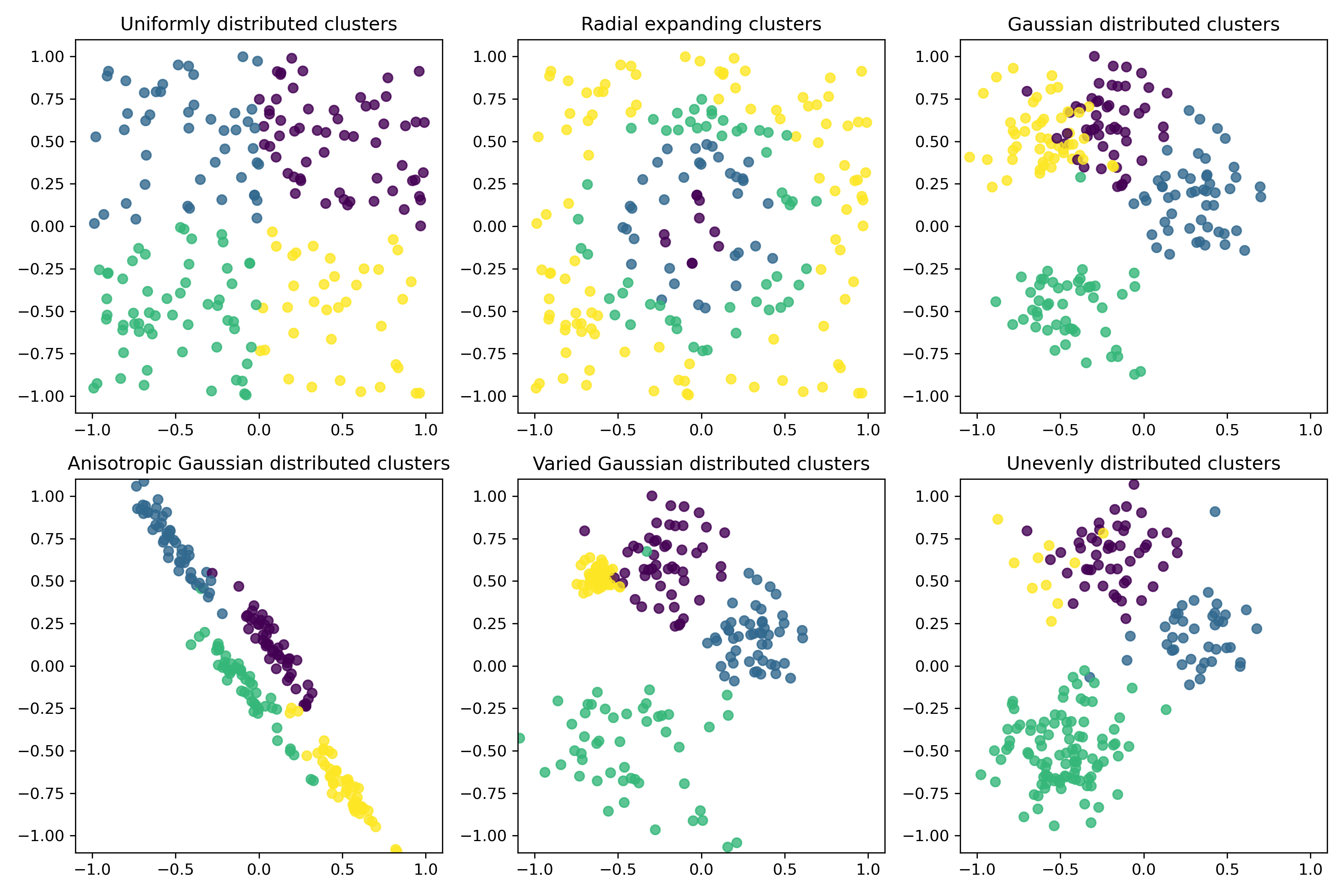}
    \caption{The examples of six scenarios}
    \label{fig:samples}
\end{figure}

To fully demonstrate the diversity of spatial distribution, we have considered the following six scenarios, with an example of each scenario shown in Figure~\ref{fig:samples}. 

\begin{itemize}
    \item[] \textbf{- Scenario 1: Uniform distributed clusters.} The points $\{s_i\}$ are uniformly sampled from $[-1,1]^2$, and the sampled points are divided into the following four groups:
    \begin{align*}
    &D_1=\{s_i  \mid s_{i1}>0, s_{i2}>0 \}, \ \ \ \ \ \ 
    D_2=\{s_i  \mid s_{i1}<0, s_{i2}>0 \}, \\
    &D_3=\{s_i  \mid s_{i1}<0, s_{i2}<0 \}, \ \ \ \ \ \ 
    D_4=\{s_i  \mid s_{i1}>0, s_{i2}<0 \}.
    \end{align*}
    \item[]\textbf{- Scenario 2: Radial expanding clusters.} The sample data are grouped based on their distance from a central point $\mu=(0,0)$, creating concentric rings of data points. The sample points are assigned to different clusters based on their distance from the center.
    \begin{align*}
    &D_1=\{s_i  \mid d(s_{i},\mu)\leq r_1 \}, \ \ \ \ \ \ 
    D_2=\{s_i  \mid r_1 < d(s_{i},\mu)\leq r_2 \}, \\
    &D_3=\{s_i  \mid r_2 < d(s_{i},\mu)\leq r_3 \}, \ \ \ \ \ \ 
    D_4=\{s_i  \mid d(s_{i},\mu)> r_3 \},
    \end{align*}    
    where $d(s_i,\mu)=\sqrt{(s_{i1}-\mu_1)^2+(s_{i2}-\mu_2)^2}$, and $r_1$, $r_2$, $r_3$ are predefined radii that determine the boundaries of the concentric groups. We set $r_1=0.25$, $r_2=0.5$, $r_3=0.75$ in our study.
    \item[] \textbf{- Scenario 3: Gaussian distributed clusters.} The sample points are divided into four domains, each centered around one of the four cluster centers $\mu_j \sim \text{Uniform}([-0.5, 0.5])$ for $j=1, 2, 3, 4$. The sample points $s_i$ within each domain are generated to follow an isotropic normal distribution centered around $\mu_j$ with a standard deviation $\sigma$. Specifically, $D_j=\{s_i  \mid  s_i \sim \mathcal{N}(\mu_j, \sigma^2),i=1,\ldots, \frac{n}{4}\}$, where $\sigma$ is set as $0.2$ in this study. 
    \item[]\textbf{- Scenario 4: Anisotropic Gaussian distributed clusters.} The sample data are initially generated as scenario 3, then a linear transformation matrix $T$ is applied to the generated data, transforming $D$ as $D=D \times T$. The transformation matrix $T$ is set as 
    \[T = \begin{bmatrix}
    0.6 & -0.6 \\
    -0.4 & 0.8
    \end{bmatrix}.\]
    \item[]\textbf{- Scenario 5: Varied Gaussian distributed clusters.} Similar to Scenario 3, the sample points are distributed around four cluster centers $\mu_j$, but with different variances $\sigma_j$. The sampled domain $D_j = \{s_i  \mid  s_i \sim \mathcal{N}(\mu_j, \sigma_{j}^2),i=1,\ldots, \frac{n}{4}\}$. In this study, we set $\sigma_1 =0.2$, $\sigma_2 =0.15$, $\sigma_3 =0.3$, and $\sigma_4 =0.06$.
    \item[]\textbf{- Scenario 6: Unevenly Gaussian distributed clusters.} Different from scenario 3, different clusters are assigned uneven numbers of samples in this scenario. The group sizes are as predefined: $D_1=50$, $D_2=40$, $D_3=100$, and $D_4=10$.
\end{itemize}

\subsection{Methods}\label{sec:3.2}
We applied the SCFA with the spatial dependency strength parameter $\phi=1$ to the simulated dataset. For the weights $w_{ij}$, we evaluated two distinct cases:
 
\begin{itemize}
    \item[] \textbf{- Five nearest neighbors.} For each location $s_i$, the five closest locations are assigned a weight of $w_{ij}=1$, otherwise $w_{ij}=0$.
   
    \item[]\textbf{- Exponential weight function} For any $s_i$ and $s_j$, weights are determined by the exponential decay of the distance:
    \begin{equation*}
        w_{ij} = \exp{(-\Vert s_i - s_j \Vert^2/0.1^2)}.
    \end{equation*}
    These two approaches are denoted by -n and -e, respectively.
\end{itemize}

As a competitor, we employed the EFA as the baseline model. Although it does not consider spatial structure, it is the most common method to identify the underlying structure of the data. By comparing the performance of both models, we can evaluate how well each approach identifies the underlying structures in the spatial data. All estimation processes in this study were carried out using the \texttt{Python} package ``FactorAnalyzer'' \citep{persson2021python}, which is based on the implementation of the corresponding \texttt{R} package ``psych'' \citep{manual2024}. This performs factor analysis, including the estimation of factor loadings, unique variances, and the common factors themselves.

To evaluate the difference between the performances of SCFA and EFA, we compare the estimated covariance matrix $\hat{A}\hat{A}^{\top}$ with the simulated covariance matrix $AA^{\top}$ because the covariance matrix effectively captures the overall contribution of the factor loadings in explaining the observed data. Note that the shapes of the loading matrices estimated by SCFA and EFA differ. The loading matrix estimated by SCFA has the shape $G\times p\times m$ while that estimated by EFA has the shape $p\times m$. This difference in shape makes a direct comparison difficult. To resolve this, we consider the loading matrices for all samples. According to the definition of EFA, all samples share the same loading matrix, resulting in a total loading matrix of the shape $n\times p \times m$. As we proposed in Section \ref{subsec:2.2}, samples within the same group share the same structure, so for a sample $x_i$ in group $g$, $A_{gi} = A_g$. This results in a loading matrix by SCFA with a shape of $n\times p \times m$. This approach ensures that the loading matrices from SCFA and EFA are comparable.

We repeated 50 experiments of data generation and fitting each method, and calculated the average value of the following metrics.
\begin{enumerate}
    \item \textbf{Frobenius Distance} This measures the magnitude of the difference between two matrices by taking the square root of the sum of the absolute squares of their elements. It is suitable for measuring overall differences between two matrices.
\[D_{\text{Frobenius}} = \frac{1}{n}\sum_{k=1}^{n} \sqrt{\sum_{i=1}^{p} \sum_{j=1}^{p} \left| (AA^{\top})_{ij}^{(k)} - (\hat{A}\hat{A}^{\top})_{ij}^{(k)} \right|^2}
\]

    \item \textbf{Wasserstein Distance} This measures the minimum ``workload'' required to transform one matrix into another matrix. It is particularly useful for comparing the overall distribution similarity of two matrices.
\[D_{\text{Wasserstein}} = \frac{1}{n}\sum_{k=1}^{n} W_2\left(AA^{\top}, \hat{A}\hat{A}^{\top}\right) = \frac{1}{n}\sum_{k=1}^{n} \inf_{\substack{ X \sim AA^{\top} \\ Y \sim \hat{A}\hat{A}^{\top} }} \ (\mathbb{E} \Vert X,Y \Vert ^2)^{1/2}\] 
   
    \item \textbf{Chebyshev Distance} This metric measures the largest absolute difference between the corresponding elements of two matrices. It focuses on the one with the largest difference in all dimensions between two matrices.
    \[D_{\text{Chebyshev}} = \max_{i,j,k} \left| (AA^{\top})_{ij}^{(k)} - (\hat{A}\hat{A}^{\top})_{ij}^{(k)} \right|\]
    
    \item \revise{\textbf{BIC} This measures a balance between the goodness of fit and the model's conciseness. A smaller BIC value suggests good prediction performance. We used criterion \eqref{IG} with $c_n = \log(n)$, ensuring an BIC-type criterion.}
\end{enumerate}

To further refine our evaluation, we considered different initial grouping methods. Specifically, we experimented with random grouping and \emph{K-means} clustering based on $\{s_i\}$, using \texttt{Python} package ``scikit-learn''~\citep{scikit-learn}. Thus, we compared the performance of EFA with various conditional SCFA, including:
\begin{itemize}
    \item \emph{K-means} initial grouping + five nearest neighbors spatial weights.
    \item \emph{K-means} initial grouping + exponential weight function.
    \item random initial grouping + five nearest neighbors spatial weights
    \item random initial grouping + exponential weight function
\end{itemize}

\subsection{Results}

\begin{table}[htp!]
\centering
\caption{Performance comparison of EFA and SCFA across six scenarios}
\label{tab:performance_comparison}
\begin{tabular}{c l c c c c}
\hline
&  & \textbf{Frobenius } & \textbf{Wasserstein} & \textbf{Chebyshev } &  \\ 
\textbf{Scenarios} & \multicolumn{1}{c}{\textbf{Models}} & \textbf{Distance} & \textbf{Distance} & \textbf{Distance} & \textbf{\revise{BIC}} \\ \hline
Uniform  & EFA & 30.66 & 8.71 & 10.87 & \revise{8452} \\ 
 & SCFA -\emph{K-means} -n & \textbf{28.59} & \textbf{7.97} & \textbf{10.59} & \textbf{\revise{3887}} \\ 
 & SCFA -\emph{K-means} -e & 29.77 & 8.39 & 10.71 & \revise{4137} \\ 
 & SCFA -random -n & 29.48 & 8.30 & 10.69 & \revise{5064} \\ 
 & SCFA -random -e & 30.06 & 8.50 & 10.75 & \revise{4957} \\ 

\hline
Radial   & EFA & 27.49 & 7.76 & 10.40 & \revise{8048} \\ 
expanding& SCFA -\emph{K-means} -n & 26.62 & 7.45 & 10.26 & \revise{4806} \\ 
 & SCFA -\emph{K-means} -e & 26.72 & 7.48 & 10.27 & \revise{4720} \\ 
 & SCFA -random -n & \textbf{26.55} & \textbf{7.43} & \textbf{10.25} & \revise{\textbf{4646}} \\ 
 & SCFA -random -e & 26.69 & 7.47 & 10.27 & \revise{4807} \\ 

\hline
Gaussian  & EFA & 30.60 & 8.69 & 11.06 & \revise{8415} \\ 
 & SCFA -\emph{K-means} -n & \textbf{29.82} & \textbf{8.42} & \textbf{10.93} & \revise{\textbf{4768}} \\ 
 & SCFA -\emph{K-means} -e & 30.05 & 8.50 & 10.95 & \revise{4801} \\ 
 & SCFA -random -n & 29.96 & 8.47 & 10.94 & \revise{5087} \\ 
 & SCFA -random -e & 30.10 & 8.52 & 10.93 & \revise{5245} \\ 

\hline
Anisotropic  & EFA & 30.27 & 8.60 & 11.23 & \revise{8315} \\ 
Gaussian  & SCFA -\emph{K-means} -n & \textbf{29.14} & \textbf{8.20} & \textbf{11.07} & \revise{\textbf{4343}} \\ 
 & SCFA -\emph{K-means} -e & 29.86 & 8.45 & 11.14 & \revise{4684} \\ 
 & SCFA -random -n & 29.29 & 8.25 & 11.09 & \revise{4493} \\ 
 & SCFA -random -e & 29.81 & 8.43 & 11.12 & \revise{4607} \\ 
 
\hline
Varied  & EFA & 30.32 & 8.63 & 11.08 & \revise{8403} \\ 
 & SCFA -\emph{K-means} -n & \textbf{29.25} & \textbf{8.25} & \textbf{10.93} & \revise{\textbf{4772}} \\ 
 & SCFA -\emph{K-means} -e & 29.67 & 8.40 & 10.94 & \revise{4831} \\ 
 & SCFA -random -n & 29.57 & 8.37 & 10.97 & \revise{5172} \\ 
 & SCFA -random -e & 29.88 & 8.48 & 10.97 & \revise{5156} \\ 

\hline
Uneven  & EFA & 29.86 & 8.50 & 10.90 & \revise{8262} \\ 
 & SCFA -\emph{K-means} -n & \textbf{28.82} & \textbf{8.12} & \textbf{10.77} & \revise{\textbf{4437}} \\ 
 & SCFA -\emph{K-means} -e & 29.09 & 8.22 & 10.81 & \revise{4530} \\ 
 & SCFA -random -n & 29.12 & 8.23 & 10.81 & \revise{4813} \\ 
 & SCFA -random -e & 29.18 & 8.25 & 13.80 & \revise{4847} \\ 

\hline
\end{tabular}%
\end{table}

The results are presented in Table~\ref{tab:performance_comparison}, which shows the performance of models in different spatial data scenarios. The SCFA is superior to EFA in all scenarios, demonstrating the effectiveness of the SCFA in spatial data. Notably, for data with clearly separated centers between groups as in Scenarios 1 and 4, the combination between \emph{K-means} initial grouping method and SCFA can effectively identify the structure of underlying variables and maximize its advantages. However, in scenarios without well-separated groups, such as radial expanding situations, obtaining appropriate initial grouping using the \emph{K-means} method may be challenging. In this case, the random method for initial grouping may better leverage the capabilities of SCFA. Also, given the limited improvement in the Chebyshev distance, a distance that focuses on the maximum absolute difference between corresponding elements of the matrix, the improvement in the SCFA must have been across the entire multidimensional space.
Furthermore, the AIC values are significantly improved. This, unlike other metrics, measures its goodness as a regression equation, assuring that it is more appropriate as a statistical model.

\begin{figure}[t]
    \includegraphics[width=1\linewidth]{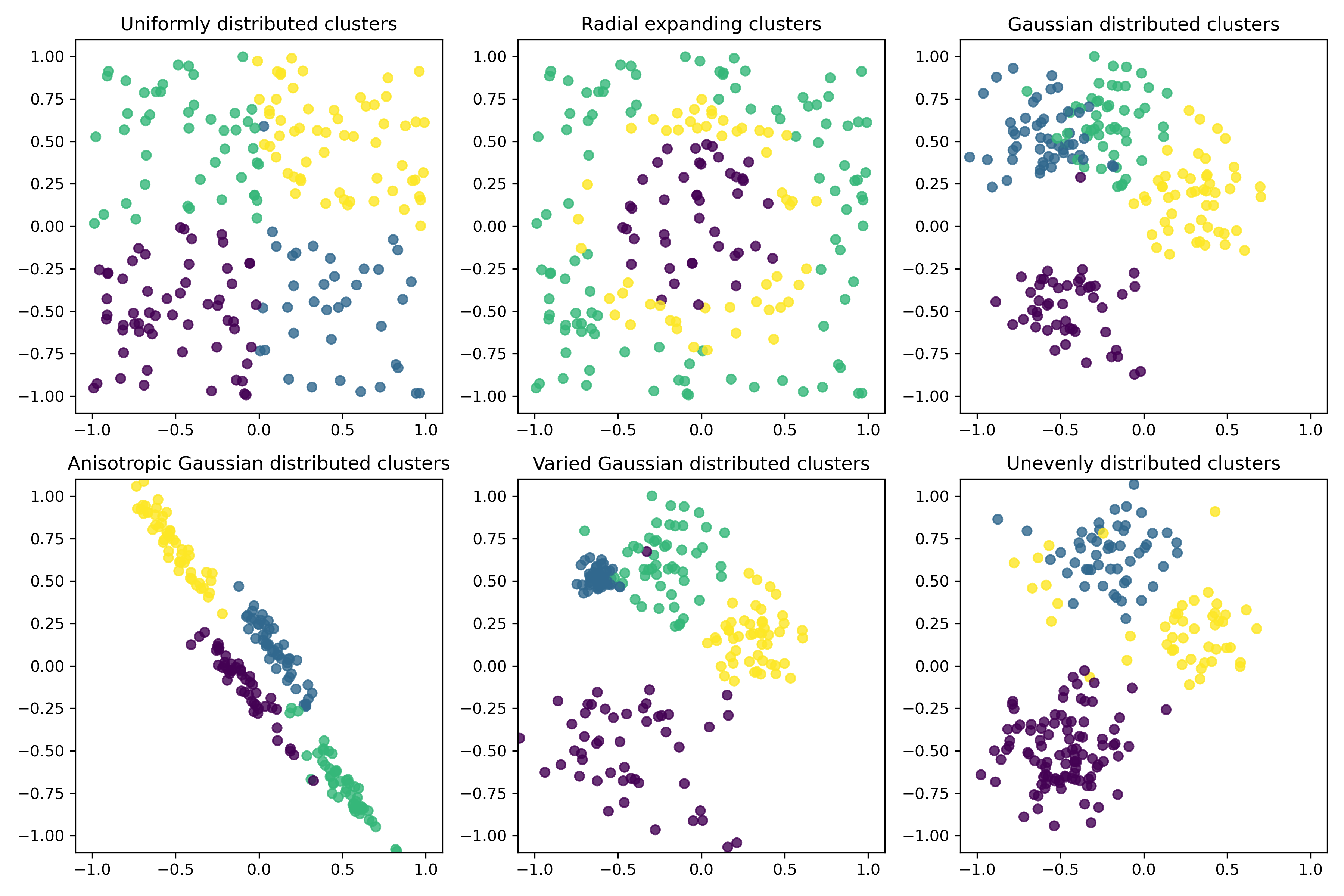}
    \caption{\revise{The grouping results obtained after applying SCFA}}
    \label{fig:SCFA_result}
\end{figure}

\revise{Figure~\ref{fig:SCFA_result} shows the grouping results after applying SCFA, demonstrating its ability to accurately capture the potential structure of the data. Due to the effective integration of factor structure and spatial information by SCFA, it is able to detect spatial heterogeneity in the structure and form clearly defined groups that reflect the original distribution patterns. However, in certain clustering configurations (such as radial expanding and unevenly distributed clustering), SCFA performs slightly weaker when some groups contain few samples. In this case, limited sample sizes pose a challenge for identifying groups because the model lacks sufficient data to capture reliable factor structures.}


\section{Application Study} \label{sec:4}
\begin{figure}[!htbp]
    \centering
    \includegraphics[width=0.6 \linewidth]{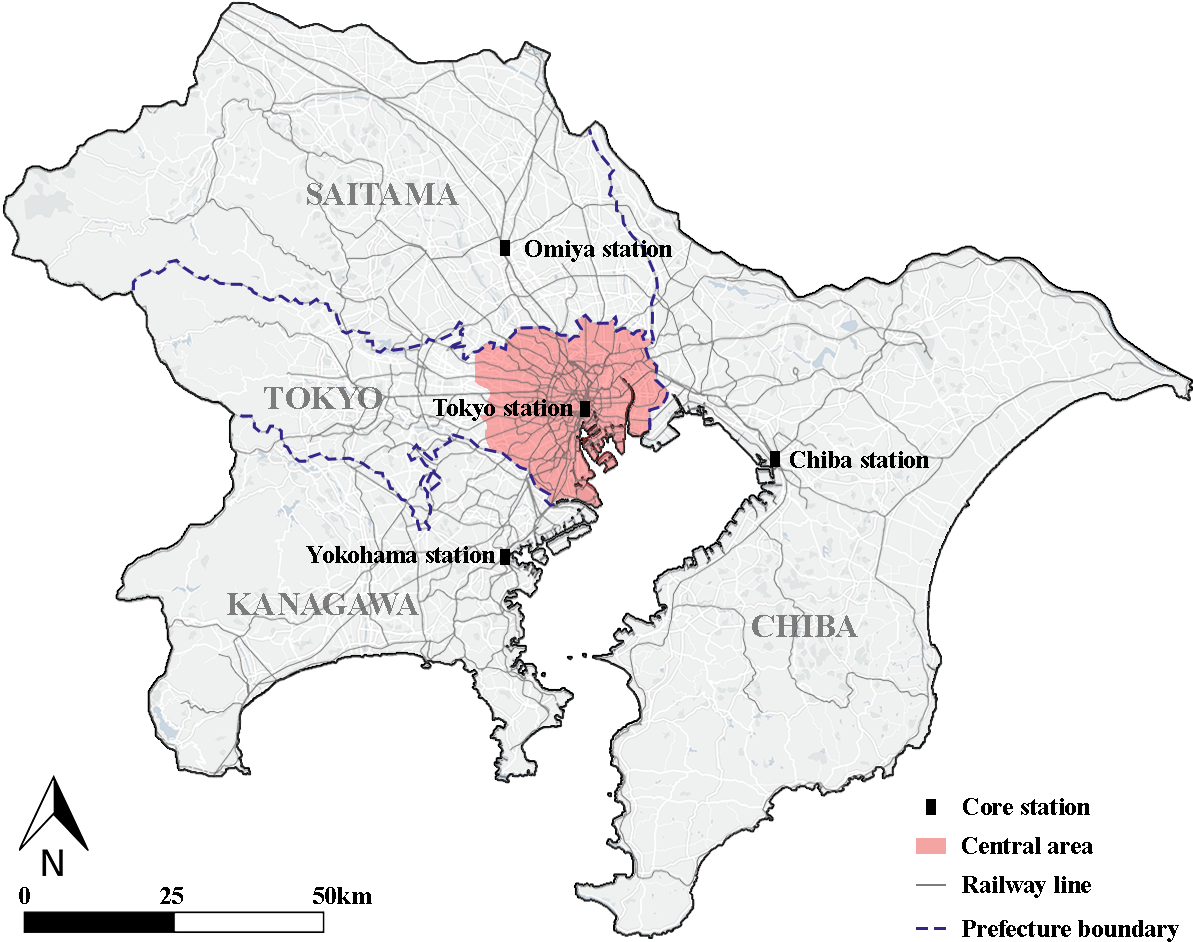}
    \caption{Tokyo metropolitan area, including Tokyo, Saitama, Chiba, and Kanagawa Prefectures.}
    \label{fig:research_area}
\end{figure}

In this section, we apply SCFA to real data. The dataset contains built environment attributes for 1535 stations in the Tokyo metropolitan area (Figure~\ref{fig:research_area}), which comes from Open Street Map\footnote{https://download.geofabrik.de/}, MLIT (Ministry of Land, Infrastructure, Transport, and Tourism)\footnote{https://nlftp.mlit.go.jp/ksj/index.html} and e-Stat\footnote{https://www.e-stat.go.jp/en} public datasets. The selection of attributes of the built environment follows our previous work \cite{jin2023understanding}, namely Residential, Employment, Commerce \& Entertainment, Transportation, and Administration \& Public attributes with five categories, each containing five indicators. Thus, for each station, we have 25-dimensional variable $X_i = (X_{i1},X_{i2},\ldots, X_{i25})$. \revise{Descriptions of the variables are provided in the appendix.}


\subsection{Experimental setting}
\begin{figure}[th]
  \centering
  \begin{tabular}{ c @{\quad} c }
    \includegraphics[width=16.2em]{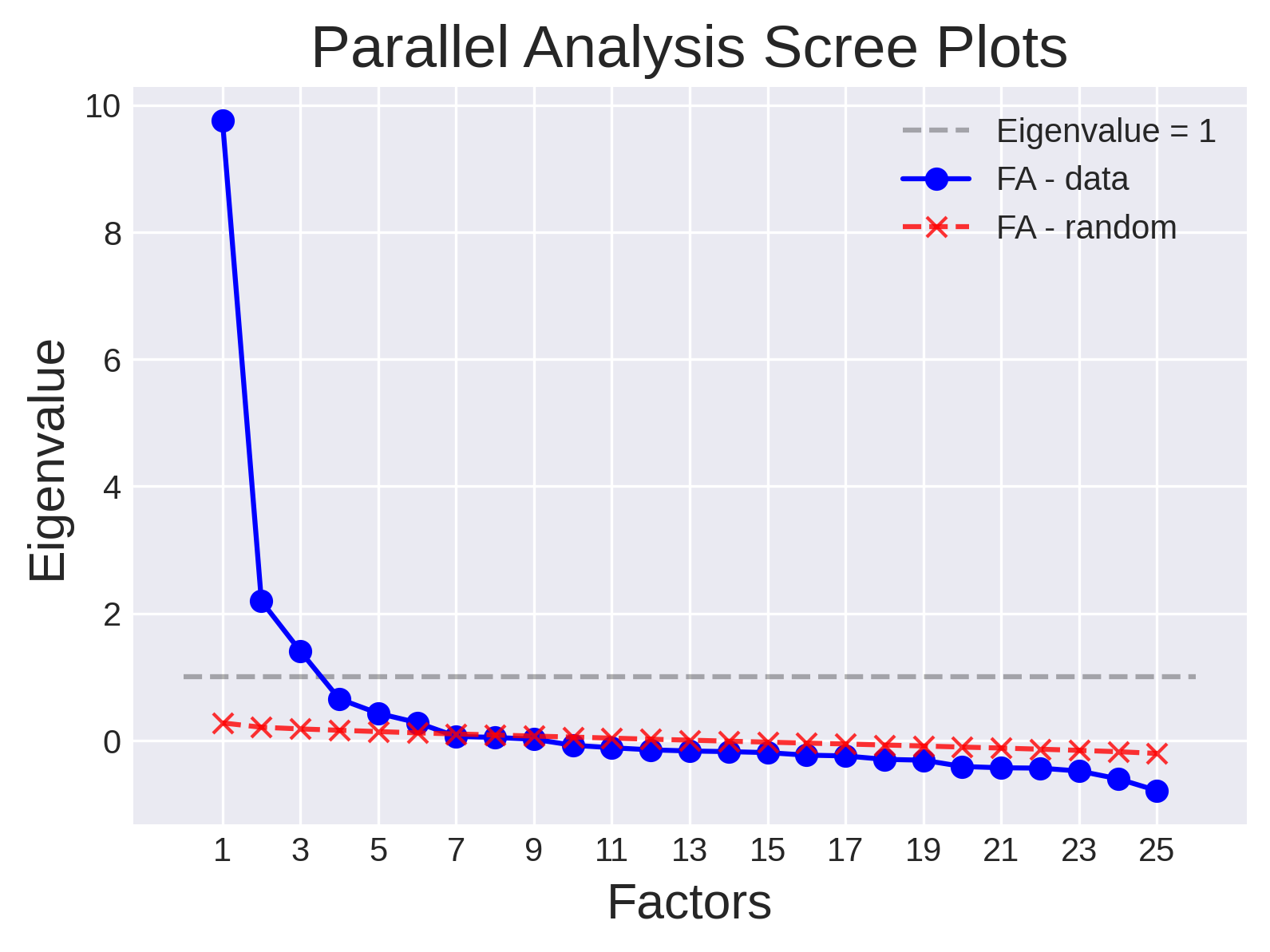} &
      \includegraphics[width=19.2em]{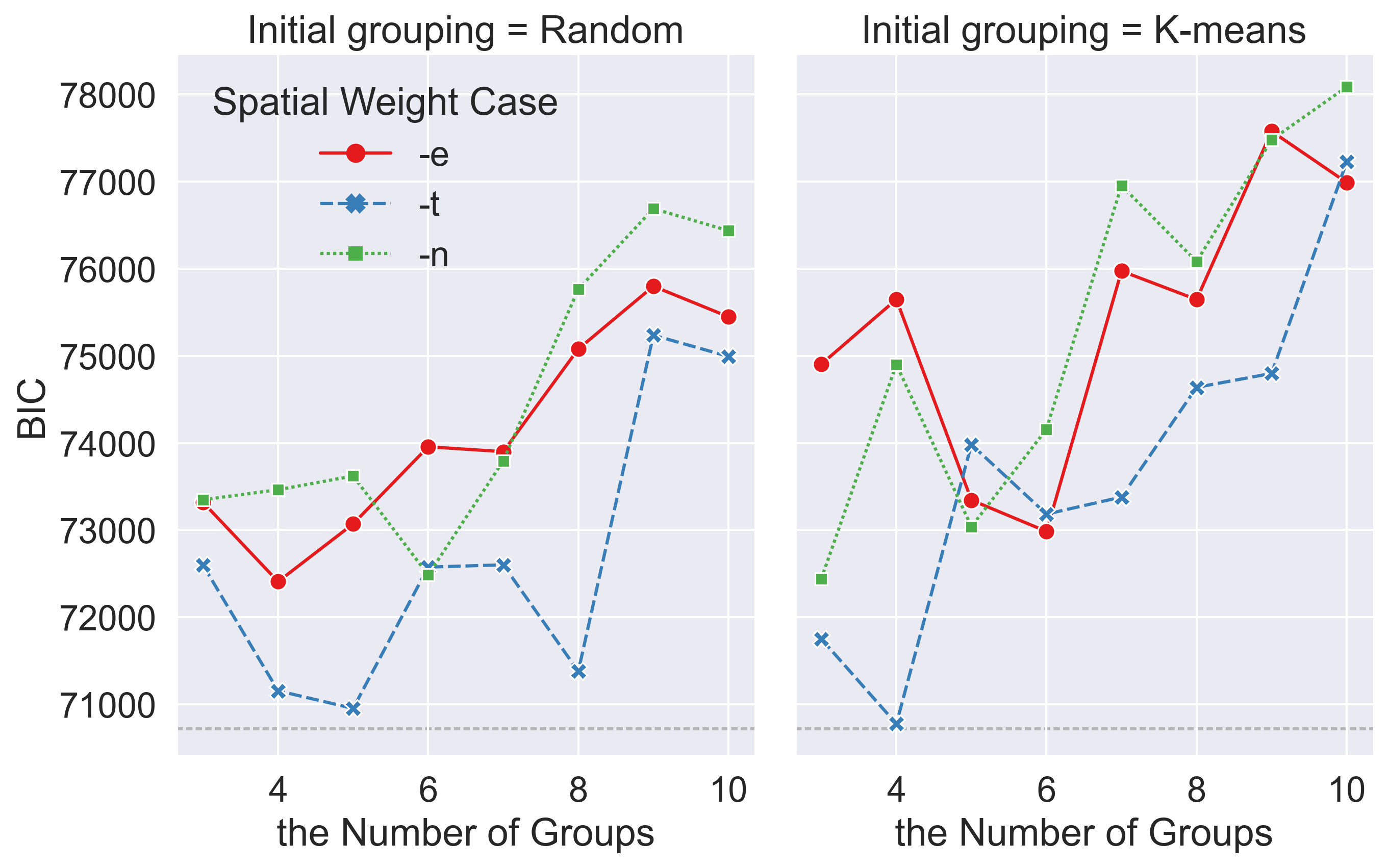}\\
    \small (a) &
      \small (b)
  \end{tabular}
  \vspace*{8pt}
  \caption{(a) Parallel analysis scree plot for selecting the number of factors. (b) BIC values as a function of the number of groups (G) for different spatial weight cases (-e, -t, -n) and two initial grouping strategies (random and \emph{K-means}). \label{fig 2}}
\end{figure}

First, it is essential to clarify the selection of three key parameters: the number of factors $m$, the spatial weight, $w$ in \eqref{penalized_likelihood_function}, and the number of groups, $G$. To determine the appropriate number of factors, we conducted a parallel analysis on the dataset. As shown in Figure~\ref{fig 2}a, when the number of factors is set to 6, the eigenvalues of the actual data exceed those of the randomly simulated datasets, ensuring the flexibility and validity of factor extraction. Regarding the spatial weight, we adopted approaches -n defined in Section~\ref{sec:3.2}. Additionally, we modified the -e method to use the Haversine distance instead of the Euclidean distance. With small differences in latitude and longitude, the small Euclidean distance makes it difficult for the model to converge. It is more appropriate to use Haversine distance that more accurately reflects real-world spatial relationships. Besides these, we also considered a case of spatial weight based on the railway's network topology, denoted as -t. Specifically, the distance between stations is defined as the shortest path distance \citep{dijkstra_note_1959}. \revise{Sensitivity analysis on parameter  $\phi$ revealed that variations within $\phi$ ranging from 0.75 to 1.25 had minimal impact on grouping results. Hence, we set $\phi =1$ and explore the number of groups $G$ from the set $G\in{1,2,\ldots,10}$. Under two initial grouping conditions, we fixed random seeds to maintain consistent initial groupings under different spatial weight configurations.} Using a BIC-type criterion \eqref{IG}, we determined that the optimal number of groups is $G = 4$, with topology-based spatial weight being the best choice. The results are illustrated in Figure~\ref{fig 2}b.



\subsection{\revise{Results of Group Identification}}
\begin{figure}[t]
    \includegraphics[width=1\linewidth]{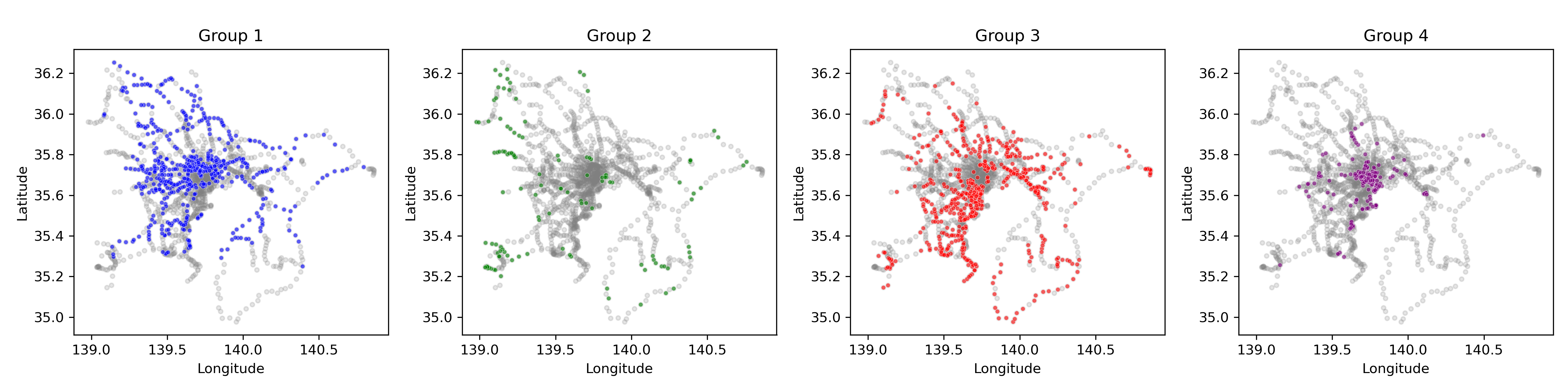}
    \caption{Station locations of four groups.}
    \label{fig:map}
\end{figure}

Figure~\ref{fig:map} shows the geographical distribution of 1535 stations divided into four groups. These results demonstrate the spatial distribution characteristics of different groups, further validating the effectiveness of SCFA in handling multivariate spatial data. The stations in Group 1 and Group 4 are mainly concentrated in the central Tokyo metropolitan area and its surrounding areas, showing a distinct spatial clustering phenomenon. In particular, Group 4 clearly captures the core area of the Tokyo metropolitan area, including Tokyo Station, Chiba Station, Yokohama Station, Omiya Station, and other key urban stations. This concentration indicates that SCFA can capture a high correlation between stations in urban core areas. Stations in core areas may share similar attributes such as high population density, high concentration of commercial activities and transportation facilities, which make these stations exhibit similar spatial dependency structures among them. In contrast, stations in Groups 2 and 3 are distributed on the periphery, reflecting a more dispersed spatial structure. The heterogeneity reflected by these peripheral stations may be related to the types of areas they serve (such as residential areas, natural landscapes or transportation junctions). SCFA successfully identifies this heterogeneity within different spatial regions and categorizes them into different groups. Furthermore, Group 3 exhibits multi-centered spatial distribution characteristics which indicate that SCFA can not only capture single-center clustering phenomena but also identify more complex patterns of spatial expansion and multi-centered clustering structures.

\subsection{Effective Underlying Structure Discovery}
\begin{figure}[t]
    \includegraphics[width=1\linewidth]{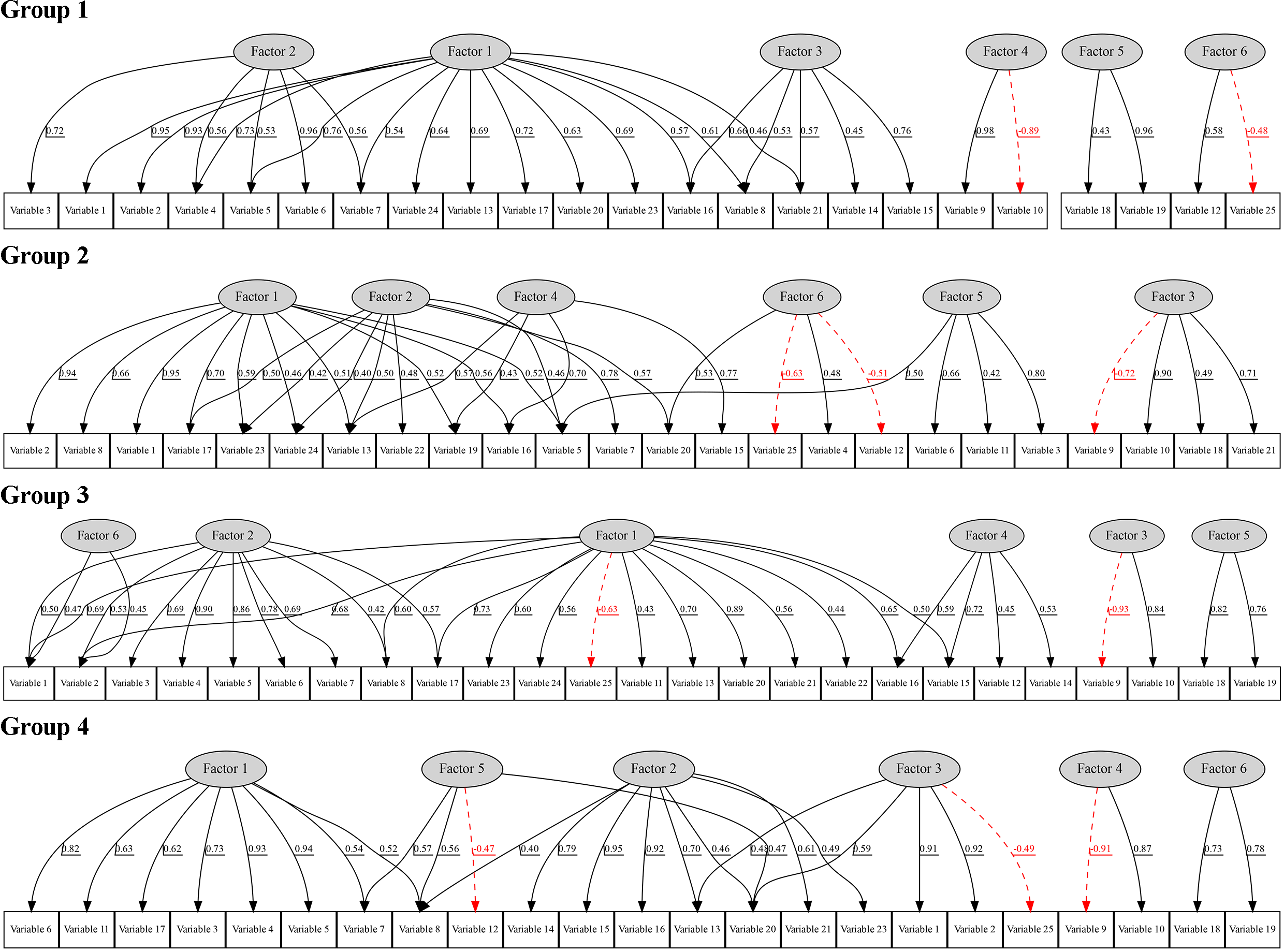}
    \caption{Underlying structures of factors among four groups}
    \label{fig:stuct}
\end{figure}

We visualized factor structure diagrams for each group and kept edges with loadings greater than 0.4 for simplicity. Figure~\ref{fig:stuct} shows the underlying structures of four groups obtained through SCFA. The factor structures of each group exhibit significant differences, clearly reflecting the impact of spatial heterogeneity on the dependency relationships for each factor model. 

\begin{itemize}

    \item[] \revise{\textbf{- Group 1.} The factor structure of Group 1 displays a relatively simple dependency pattern, dominated by Factor 1 and Factor 2. Factor 1 primarily influences residential attributes,  such as resident density (Variable 1), household density (Variable 2), and walkability (Variable 5), indicating that this factor represents urban characteristics related to high-density convenient residential areas. Factor 2 reflects employment and housing price attributes, including worker density (Variable 6), company density (Variable 7), and housing price (Variable 3). This suggests that labor resources and housing market dynamics play a significant role in this group.}

    \item[]\revise{\textbf{- Group 2.} This group presents a more complex structure, with more variables for Factors 1, 2, and 4, reflecting more diverse spatial characteristics. While Factor 1 and Factor 2 again focus on residential attributes, they also have strong loadings on commercial (e.g., number of restaurants, Variable 13) and transportation variables (e.g., transfer lines, Variable 17; intersections, Variable 19; bus stops, Variable 20). This combination likely corresponds to suburban areas with a high dependency on transportation and localized business services, where residential and commercial functions coexist.}
    \item[] \revise{\textbf{- Group 3.} Factor 1 of Group 3 influences a wider range of variables, especially those related to population density, transportation, and commercial attributes. This suggests that Group 3 may correspond to spatial regions with diversified functional uses. Factor 2 has strong loading in both employment and residential aspects, reflecting the mixed-use of the area where living areas and work areas are highly integrated.} 
    \item[]\revise{\textbf{- Group 4.} The factor structure of Group 4 is relatively simpler, with Factor 1 exerting a strong influence on employment and commercial characteristics, while Factor 2 predominantly affects transportation and commercial variables. This pattern suggests that Group 4 represents the area with a high concentration of commercial activities, potentially indicative of a central business district or a densely developed commercial area.}"
\end{itemize}

Furthermore, there are similarities in the underlying structures among the four groups. For example, Factor 1 dominates the dependencies of most variables in each group; Variable 9 \revise{(IT sector proportion)} and Variable 10 \revise{(major company proportion)} often jointly exert strong influences on the same factor. This reveals some global characteristics in the underlying structures, where even with spatial heterogeneity present, different regions still share certain common patterns. On the one hand, Factor 1 serves as a global dominant factor that continues to govern dependencies among numerous variables in each group. This may suggest that certain global variables (such as infrastructure, population density, and economic activities) exhibit commonalities across different regions within the Tokyo metropolitan area. These variables might similarly affect dependencies between various regions, leading to the importance of factor 1 within all groups. On the other hand, whether in urban central areas or peripheral regions, \revise{maintaining consistent relationships between Variables 9 and Variable 10 points to potential economic interdependence. High proportions of IT departments and major companies typically indicate that an area is a financial hub where technology and business headquarters drive overall economic vitality. This coordination transcends spatial boundaries to form uniform interactions.} 

Finally, we compared the performance of EFA and SCFA on real data using AIC and BIC as evaluation criteria. The results, as presented in Table~\ref{tab: fitting performance}, show that the SCFA outperforms the EFA in both criteria. This underscores that spatial grouping in the SCFA works effectively in real data.

\setlength{\tabcolsep}{30pt}
\begin{table}[htp!]
\centering
\caption{Comparison of AIC and BIC between EFA and SCFA Models}
\label{tab: fitting performance}
\begin{tabular}{ c c c }
\hline
\textbf{Models} & {\textbf{AIC}} & \textbf{BIC} \\ 
\hline
EFA & 75462 & 76396 \\ 
SCFA & 66425 & 70160  \\ \hline
\end{tabular}%
\end{table}

\section{Conclusion} \label{sec:5}
In this study, we developed a novel spatially clustered factor analysis (SCFA) that aims to uncover the underlying dependency structures in multivariate spatial data, addressing the challenges of spatial heterogeneity and complexity in spatial data. Our approach integrates the clustering method with exploratory factor analysis, allowing us to partition the data into clusters where locations within the same group share identical dependency structures. This combination enhances the ability to capture underlying structural differences caused by spatial heterogeneity, which traditional exploratory factor analysis methods might overlook. Through comprehensive simulation studies and empirical applications, we have confirmed that SCFA can effectively capture structural differences caused by spatial heterogeneity in various spatial distributions, leading to more accurate variable relationships and factor structures.

In addition, SCFA can effectively identify spatial dependency relationships within different locations as well as global dependencies. This ability is crucial for analyzing intricate data across regions and scales. By capturing these diverse dependency structures, SCFA enables researchers to gain deeper insights into interactions among different regions in complex systems and their influence on the overall structure. Whether in policy making, resource allocation, or strategic planning, accurately understanding multilevel dependency relationships improves the precision and efficiency of decision-making. \revise{While our method demonstrates improved performance on spatially heterogeneous data, it is not intended for non-spatial datasets, as it relies on spatial attributes to capture meaningful patterns. The applicability and superiority of this approach are observed primarily when the data exhibits spatial heterogeneity. Future work could further investigate the theoretical conditions under which this method excels, providing a more rigorous mathematical foundation for its use.}

\section*{Source Code }
The source code for Sections \ref{sec:3} and \ref{sec:4} is available at the GitHub repository (\url{https://github.com/yanxiuJin/Spatially_Clustered_Factor_Analysis}).

\section*{Acknowledgments}
This work was supported by JST SPRING, Grant Number JPMJSP2108, and JSPS KAKENHI Grant Numbers 22J21090, 21H00699 and 24KJ0750.

\section*{Appendix}

\renewcommand{\thetable}{A\arabic{table}}
\renewcommand{\thefigure}{A\arabic{subsection}-\arabic{figure}}
\renewcommand{\thesubsection}{A\arabic{subsection}}
\setcounter{table}{0} 
\setcounter{section}{1}
\setcounter{figure}{0}

\revise{The station environment data used in this study is classified into five categories based on functional attributes: residential, employment, commercial and entertainment, transportation, and administrative and public attributes. According to the variables identified in previous studies \citep{jin2023understanding}, five variables were selected for each category. Table~\ref{tab:Variables} presents the range of independent variables and their sources.}

\setlength{\tabcolsep}{5pt} 

\renewcommand{\arraystretch}{1.25} 
\begin{table}[!h]
\centering

\footnotesize
\caption{\revise{Variables of the station built environment.}}
\label{tab:Variables}
\begin{tabular}{c l c c c c}
\hline

\textbf{Category} & \multicolumn{1}{c}{\textbf{Variable}} & \textbf{Scope}& \textbf{Data Source} \\ \hline
Residential & Variable 1 - resident density & 1.5 km & MLIT \\
attributes & Variable 2 - household density &  1.5 km & e-Stat \\
& Variable 3 - housing price  & 1.5 km & MLIT  \\
& Variable 4 - resident consumption & 1.5 km & e-Stat  \\
& Variable 5 - walkability index \(^1\) & – & Real Estate Homepage\(^2\) \\
\hline
Employment & Variable 6 - worker density & 800 m & e-Stat \\
attributes & Variable 7 - company density &  800 m & e-Stat  \\
& Variable 8 - financial sector proportion & 800 m & e-Stat  \\
& Variable 9 - IT sector proportion & 800 m & e-Stat  \\
& Variable 10 - major company proportion & 800 m & e-Stat  \\
\hline
Commerce and & Variable 11 - number of shopping malls & 800 m & OSM \\
entertainment & Variable 12 - number of restaurants &  800 m & OSM\\
attributes & Variable 13 - number of entertainments &  800 m & OSM\\
& Variable 14 - number of retails & 800 m & OSM  \\
& Variable 15 - commercial area & 800 m & MLIT  \\
\hline
Transportation & Variable 16 - number of transfer lines & Station & Homepage of JR station \\
attributes & Variable 17 - passenger load & Station & Homepage of JR station  \\
& Variable 18 - number of intersections & 800 m & OSM \\
& Variable 19 - number of bicycle parking & 800 m & OSM \\
& Variable 20 - number of bus stops & 800 m & OSM  \\
\hline
Administration & Variable 21 - green park area & 800 m & MLIT \\
and public & Variable 22 - number of administration facilities & 800 m & OSM \\
attributes & Variable 23 - number of public facilities & 800 m & OSM  \\
& Variable 24 - number of education facilities & 800 m & OSM \\
& Variable 25 - number of sports centers & 800 m & MLIT  \\
\hline

\end{tabular}%

\small{
\begin{tabbing}
\(^1\)The walkability index is an indicator that assesses the convenience of living within walking distance. \\
\(^2\) \url{https://lifullhomes-index.jp/}
\end{tabbing}
}
\end{table}

\clearpage
\newpage
\bibliographystyle{chicago}
\bibliography{ref}

\end{document}